\documentclass[fleqn,12pt,twoside]{article}
\usepackage{espcrc1}

\usepackage{graphicx}
\usepackage[figuresright]{rotating}

\makeatletter
\def\simleq{\mathrel{\mathpalette\gl@align<}}
\def\simgeq{\mathrel{\mathpalette\gl@align>}}
\def\gl@align#1#2{\lower.6ex\vbox{\baselineskip\z@skip\lineskip\z@
     \ialign{$\m@th#1\hfill##\hfil$\crcr#2\crcr\sim\crcr}}}
\makeatother
\newcommand{\bec}[1]{\mbox{\boldmath $#1$}}

\newcommand{\bra}{\langle}
\newcommand{\ket}{\rangle}
\newcommand{\braket}[1]{\bra #1 \ket}
\newcommand{\qq}{\braket{\bar{q}q}}

\newcommand{\qGq}{g\braket{\bar{q}\sigma_{\mu\nu}G_{\mu\nu} q}}
\newcommand{\AmS}{{\protect\the\textfont2
  A\kern-.1667em\lower.5ex\hbox{M}\kern-.125emS}}

\title{Quark-Gluon Mixed Condensate 
$g\langle\bar{q}\sigma_{\mu\nu}G_{\mu\nu} q\rangle$ from Lattice QCD}

\author{Takumi Doi\address[titech]{Dept. of Phys., 
			Tokyo Institute of Technology, 
			Meguro, Tokyo 152-8551, Japan},
	\thanks{E-mail: doi@th.phys.titech.ac.jp}
	Noriyoshi Ishii\address[riken]{Radiation Laboratory, RIKEN, 
			Hirosawa 2-1, Wako, Saitama, 351-0198, Japan},
	Makoto Oka\addressmark[titech]
	and
	Hideo Suganuma\addressmark[titech]}
       
\begin{document}

% typeset front matter
\maketitle

%%%%%%%%%%%%%%%%%%%%%%%%%%%%%%%%%%%%%%%%%%%%%%%%%%%%%%%%%%%%%%%%%
%%%%%%%%%%%%%%%%%%%%%%%%%%%%%%%%%%%%%%%%%%%%%%%%%%%%%%%%%%%%%%%%%
\begin{abstract}

We study the quark-gluon mixed condensate 
$g\langle\bar{q}\sigma_{\mu\nu}G_{\mu\nu} q\rangle$, which is another chiral order parameter,  
in SU(3)$_c$ lattice QCD with the Kogut-Susskind fermion at the quenched level.
Using 100 gauge configurations on the $16^4$ lattice with $\beta = 6.0$, 
we measure $g\langle\bar{q}\sigma_{\mu\nu}G_{\mu\nu} q\rangle$ 
at 16 points in each gauge configuration for each current-quark mass of $m_q=21, 36, 52$ MeV.
From the 1600 data for each $m_q$, 
we find $m_0^2 \equiv 
g\langle\bar{q}\sigma_{\mu\nu}G_{\mu\nu} q\rangle / 
\langle\bar{q}q\rangle
\simeq 2.5$ GeV$^2$ 
at the lattice scale of $a^{-1} \simeq 2 {\rm GeV}$ in the chiral limit.
The large value of 
$g\langle\bar{q}\sigma_{\mu\nu}G_{\mu\nu} q\rangle$
suggests its importance in the operator product expansions in QCD.
We also show our preliminary results of $\qGq$ at finite temperature.
\end{abstract}

%%%%%%%%%%%%%%%%%%%%%%%%%%%%%%%%%%%%%%%%%%%%%%%%%%%%%%%%%%%%%%%%%
%%%%%%%%%%%%%%%%%%%%%%%%%%%%%%%%%%%%%%%%%%%%%%%%%%%%%%%%%%%%%%%%%
\section{The importance of the quark-gluon mixed condensate $\qGq$}
\label{sec:intro}

In order to understand the non-perturbative structure
of the QCD vacuum, it is important to study 
various condensates such as $\qq$ and 
$\langle G_{\mu\nu}G^{\mu\nu} \rangle$.
Among various condensates, we emphasize the importance 
of the quark-gluon mixed condensate 
$\qGq \equiv 
{g\braket{\bar{q}\sigma_{\mu\nu}G_{\mu\nu}^A \frac{1}{2}\lambda^A q}}$.
First, the mixed condensate represents a direct correlation 
between quarks and gluons in the QCD vacuum. 
In this point, the mixed condensate differs from $\qq$ and 
$\langle G_{\mu\nu}G^{\mu\nu} \rangle$ even at the qualitative level.
Second, this mixed condensate is another chiral order parameter 
of the second lowest dimension, because the chirality of the quark flips as
$
\qGq = g\braket{\bar{q}_R\ (\sigma_{\mu\nu}G_{\mu\nu})\ q_L}
+ g\braket{\bar{q}_L\ (\sigma_{\mu\nu}G_{\mu\nu})\ q_R}.
$
Third, the mixed condensate plays an important role in
various QCD sum rules, especially in the baryons\cite{Ioffe,Dosch},
the light-heavy mesons~\cite{Dosch2}
and the exotic mesons~\cite{Latorre}.
In the QCD sum rules,
the value $m_0^2 \equiv \qGq / \qq \simeq 0.8\pm 0.2\ {\rm GeV}^2$ has been 
proposed as a result of the 
phenomenological analyses~\cite{Bel}.
However, in spite of the importance of $\qGq$, 
there was only one preliminary
lattice QCD work~\cite{K&S}, 
which was performed with very little 
statistics (only 5 data)
using a small ($8^4$) 
and coarse lattice ($\beta=5.7$).

Therefore, we present the calculation for $\qGq$ in 
lattice QCD with a larger $(16^4)$ and finer $(\beta=6.0)$
lattice and with high statistics (1600 data).
We perform the measurement of $\qGq$ as well as $\qq$
in the SU(3)$_c$ lattice at the quenched level, 
using the Kogut-Susskind (KS) fermion to keep the chiral symmetry.
We generate 100 gauge configurations and pick up 16 points 
for each configuration to calculate the condensates.
With this high statistics of 1600 data for each quark mass,
we perform reliable estimate for 
the ratio $m_0^2\equiv \qGq/\qq$ at the lattice scale 
in the chiral limit~\cite{DOIS:qGq}.

%%%%%%%%%%%%%%%%%%%%%%%%%%%%%%%%%%%%%%%%%%%%%%%%%%%%%%%%%%%%%%%%%
%%%%%%%%%%%%%%%%%%%%%%%%%%%%%%%%%%%%%%%%%%%%%%%%%%%%%%%%%%%%%%%%%
\section{Lattice Formalism}
\label{sec:formalism}

We first emphasize that
both of the  condensates  $\qq$  and  $\qGq$
work as the chiral order parameter, and therefore
to keep chiral symmetry is essential for our study.
From this viewpoint, we adopt the KS-fermion, which can 
preserve the explicit chiral symmetry for the quark mass $m=0$, unlike 
the Wilson and  the clover fermions.

The action of the KS-fermion is described by 
spinless Grassmann fields $\bar{\chi},\chi$
and the gauge link-variable $U_\mu \equiv \exp[ -iagA_\mu ]$.
In the absence of the gauge field, 
the SU(4)$_f$ quark-spinor field $q$ with spinor $i$ and flavor $f$
is expressed 
by $\chi$ as 
\begin{eqnarray}
\label{eq:q-ks_trans}
q_i^f (x) 
&=&  \frac{1}{8}
\sum_{\rho}\
( \Gamma_\rho )_{if}\ 
\chi (x+\rho ), \ \ 
\Gamma_\rho \equiv \gamma_1^{\rho_1} \gamma_2^{\rho_2} \gamma_3^{\rho_3} \gamma_4^{\rho_4}, \ \ \rho \equiv (\rho_1,\rho_2,\rho_3,\rho_4)
\end{eqnarray}
where $\rho$ with $\rho_\mu \in \{0,1\}$ runs over the 16 sites 
in the $2^4$ hypercube.
% and $i, f$ denote the spinor and the flavor indices, respectively.
When the gluon field is turned on, 
we insert
additional link-variables in Eq.~(\ref{eq:q-ks_trans})
to respect the gauge covariance.
Hence, the flavor-averaged condensates are expressed as 
\begin{eqnarray}
\label{eq:condensates-qq-def} 
%&&
%\lefteqn{
&&a^3 \qq 
= - \frac{1}{4}\sum_f {\rm Tr}\left[ \braket{q^f(x) \bar{q}^f(x)} \right] 
%%%\right]_{\stackrel
%%%	{\mbox{$\scriptstyle \mbox{\tiny color}$}}
%%%	{\mbox{$\scriptstyle \mbox{\tiny spinor}$}}}
%
%
= - \frac{1}{2^8}\sum_\rho
       {\rm Tr}\left[ \Gamma_\rho \Gamma_\rho^\dag\ 
	\braket{\chi(x+\rho ) \bar{\chi}(x+\rho )} \right], \\
\label{eq:condensates-qGq-def} 
%&&
%\lefteqn{
&&a^5 \qGq
= - \frac{1}{4}\sum_f \sum_{\mu,\nu}{\rm Tr}
	\left[ \braket{q^f(x) \bar{q}^f(x)} \sigma_{\mu\nu} G_{\mu\nu}\right]  \nonumber \\
%
%&& 
%\qquad 
&&= - \frac{1}{2^8} \sum_{\mu,\nu} \sum_\rho
{\rm Tr}\left[\
  {\cal U}_{\pm\mu,\pm\nu}(x+\rho)\ 
\Gamma_{\rho'} \Gamma_{\rho}^\dag\
\braket{\chi (x+\rho') 
\bar{\chi}(x+\rho)} \
\sigma_{\mu\nu}\ 
  G_{\mu\nu}^{\rm lat}(x+\rho)\ 
\right],
\end{eqnarray}
where $\rho'$ is defined as  $\rho' \equiv \rho \pm \mu \pm \nu$, 
and the sign $\pm$ is taken such that the sink point $(x+\rho')$
belongs to the same hypercube of the source point $(x+\rho)$.
Here, 
${\cal U}_{\mu,\nu}(x) \equiv 
	\frac{1}{2}\left[\ U_\mu (x) U_\nu (x+\mu ) + U_\nu (x) U_\mu (x+\nu )\ \right]$ 
is introduced to keep the gauge covariance.
%
%Here, the trace ``${\rm  Tr}$'' refers to  the spinor and  the color
%indices,  and $\langle \rangle$  denotes the  Euclidean propagator.
%
%
%
%We also introduce ${\cal U}_{\mu,\nu}$ to respect the gauge covariance,
%defined as
%
%
%\begin{eqnarray}
%{\cal U}_{\mu,\nu}(x) \equiv 
%	\frac{1}{2}\left[\ U_\mu (x) U_\nu (x+\mu ) + U_\nu (x) U_\mu (x+\nu )\ \right].
%
%\label{eq:qGq-def-U}
%\end{eqnarray}
%
%
%
%%%%%%%%%%%%%%%%%%%%%%%%%%
%
%
We adopt the clover-type definition of
the gluon field strength $G_{\mu\nu}$ on the lattice as 
\begin{eqnarray}
\label{eq:clover}
G_{\mu\nu}^{\rm lat}(s) &=& \frac{i}{16} \sum_A 
\sum_{\stackrel
	{\mbox{$\scriptstyle s'= s,s-\mu,$}}
	{\mbox{$\scriptstyle \ \ \ \ \ s-\nu,s-\mu-\nu$}}}
\lambda^A \ {\rm Tr}
\left[\ 
 \lambda^A\{ U_{\mu\nu}(s') - U_{\nu\mu}(s') \} 
\ \right],
%
%%% & {\displaystyle \mathop{\longrightarrow}_{a\rightarrow 0}} & 
%%% a^2 \left[\ g G^A_{\mu\nu}(s) \frac{\lambda^A}{2} + {\cal O}(a^2)\ \right].
%
\end{eqnarray}
which has no ${\cal O}(a)$ discretization error.
(This benefit is not available in Ref.~\cite{K&S}.)
%and thus our results suffer from less systematic errors.

%%%We show the diagrams corresponding to 
%%%Eqs.~(\ref{eq:condensates-qq-def}),(\ref{eq:condensates-qGq-def}),(\ref{eq:clover}) 
%%%in figure~\ref{fig:diags}.
%%%
%%%
%%%\begin{figure}[tbh]
%%%\begin{center}
%%%\includegraphics[scale=0.8]{qq_qGq.diagrams.eps}
%%%\caption{\label{fig:diags}
%%%The diagrammatic representation of $\qq$ (left diagram) and
%%%$\qGq$ (middle and right diagrams).
%%%The solid line with the arrow denotes the propagator of $\chi$, and
%%%the wavy line the gauge link.
%%%}
%%%\end{center}
%%%\end{figure}

%%%%%%%%%%%%%%%%%%%%%%%%%%%%%%%%%%%%%%%%%%%%%%%%%%%%%%%%%%%%%%%%%
%%%%%%%%%%%%%%%%%%%%%%%%%%%%%%%%%%%%%%%%%%%%%%%%%%%%%%%%%%%%%%%%%
\section{The lattice QCD results}
\label{sec:results}

We calculate the condensates $\qq$ and $\qGq$ using the SU(3)$_c$ lattice QCD at the quenched level.
We use the standard Wilson action at $\beta=6.0$ 
on the $16^4$ lattice.
The lattice unit $a\simeq 0.10{\rm fm}$ is obtained so as to reproduce 
the string tension $\sigma = 0.89 {\rm GeV/fm}$~\cite{Takahashi}.
We use the quark mass $m = 21, 36, 52$ MeV
(i.e. $ma = 0.0105,\ 0.0184,\ 0.0263$).
For the fields $\chi$, $\bar{\chi}$, the anti-periodic condition 
is imposed.
%The dependence on the boundary condition will be discussed later.
%
%
%
%
%We take the source point $(x+\rho)$ in 
%Eqs.~(\ref{eq:condensates-qq-def}),(\ref{eq:condensates-qGq-def}) as follows.
We measure the condensates 
on 16 different space-time points $x$ in each
configuration as $x=(x_1,x_2,x_3,x_4)$ with $x_\mu \in \{0, 8\}$ 
in $\bec{R}^4$ in the lattice unit.
%Note that $\rho$ is taken such that $\rho$ runs over the all sites 
%in the hypercube for each given $x$.
%
%
For each $m$, we calculate the flavor-averaged condensates,
and average them over the 16 space-time points and 
100 gauge configurations.
%Statistical errors are calculated using the jackknife error estimate.

\begin{figure}
%\begin{minipage}[t]{155mm}
%\begin{center}
\includegraphics[scale=0.6]{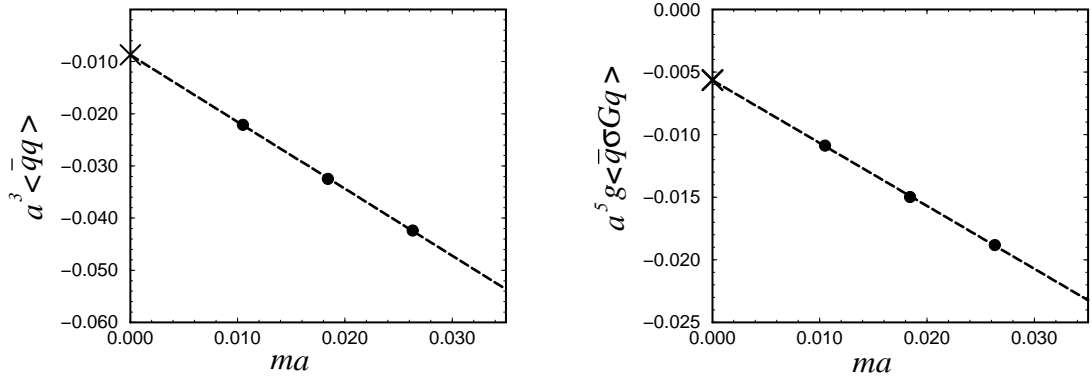}
\vspace*{-8mm}
\caption{\label{fig:plot}
The bare condensates $\qq$ and $\qGq$ plotted against the quark mass $ma$.
%at $\beta=6.0$.
The dashed lines denote the best linear extrapolations, 
and the cross symbols correspond to the values in the chiral limit. 
The jackknife errors are hidden in the circles.
}
\vspace*{-3mm}
%\end{center}
%\end{minipage}
%\hspace{\fill}
\end{figure}

%%%%%%%%%%%%%%%%%%%

%\vspace*{-3mm}

\begin{table}[thb]
%\begin{center}
\caption{
The numerical results of $\qq$ and $\qGq$ for various $ma$.
The last column denotes their values in the chiral limit by the chiral extrapolation.
\label{tab:mass-beta-6.0}
}
\begin{tabular}{ccccc}
\hline
	   &  $ma=0.0263$     &  $ma=0.0184$     &  $ma=0.0105$      &  chiral limit\\
\hline
$a^3\qq$   &  $-0.042397(16)$ &  $-0.032470(15)$ &  $-0.022124(16)$  & 
$-0.008721(17)$ \\
$a^5\qGq$  &  $-0.018820(15)$ &  $-0.014979(14)$ &  $-0.010884(14)$  & 
$-0.005652(14)$ \\
\hline
\end{tabular}
%\end{center}
\end{table}

%\vspace*{-5mm}

Figure~\ref{fig:plot} shows the 
bare condensates $\qq$ and $\qGq$ against the quark mass $ma$.
We emphasize that the jackknife errors are almost negligible, 
due to the high statistics of $1600$ data for each quark mass $m$.
Since both of $\qq$ and $\qGq$ show 
a clear linear response to $m$,
we fit the data  linearly and determine 
the condensates in the chiral limit.
The obtained data are summarized in Table~\ref{tab:mass-beta-6.0}.
%
%
%%%%%%%%%%%%%%%%%%%%%%%%%%%%%%%%%%%%%%%%%%%%%%%%%%%%%%%%%%
%
%
%To check the reliability of our results,
To check the finite volume artifact,
%we consider the finite volume artifact.
%As indicated by the Banks-Casher formula,
%the total volume $V$ should be large enough before the quark mass goes to zero.
we calculate the condensates imposing the periodic boundary 
condition on $\chi, \bar{\chi}$,
instead of the anti-periodic condition as before.
%If the physical volume is not large enough,
%the results would change.
% by using the different boundary condition.
The results with different boundary conditions almost coincide within 
about 1\% deviation.
%This suggests dependence on the boundary condition is negligible.
Therefore, we conclude that the physical volume
$V \sim (1.6\ {\rm fm})^4$ in our simulations is large enough to 
avoid the finite volume artifact~\cite{DOIS:qGq}.

%%%%%%%%%%%%%%%%%%%%%%%%%%%%%%%%%%%%%%%%%%%%%%%%%%%%%%%%%%%%%%%%%
%%%%%%%%%%%%%%%%%%%%%%%%%%%%%%%%%%%%%%%%%%%%%%%%%%%%%%%%%%%%%%%%%

%\subsection{Determination of $m_0^2=\qGq/\qq$}
%\label{sec:M0}

The values  of the  condensates in the  continuum limit  
are to be obtained through the renormalization,
which, however, suffers from
uncertainty of the non-perturbative effect.
As a more reliable quantity,
we provide   the  ratio  $m_0^2  \equiv  \qGq   /  \qq$,
which is free  from the  uncertainty from  the wave
function  renormalization  of the  quark.
%
%
%In  addition, the  dependence of  $m_0^2$  on the  lattice spacing  is
%reduced  to $a^2$,  while  $\qq$, $\qGq$  are  proportional to  $a^3$,
%$a^5$, respectively.
%
%
%We also note  that $m_0^2$ itself represents the importance 
%of $\qGq$ in OPE.
%because $\qGq$ appears as the next-to-leading 
%term to $\qq$ in  OPE in usual QCD sum rules and 
%can be parameterized by $m_0^2$.
%
%without referring to the absolute  value.

Now, we present the result of $m_0^2$ using our bare results
of SU(3)$_c$ lattice QCD as
%
%
%
%
%We find
%\vspace*{-2mm}
\begin{eqnarray}
m_0^2 \equiv \qGq / \qq  \simeq  2.5\ {\rm GeV}^2 \qquad (\beta = 6.0 \ \ {\rm or} \ \ a^{-1} \simeq 2{\rm GeV}).
%\vspace*{-2mm}
\end{eqnarray}
We see that $m_0^2$ is  rather large, which suggests the importance of
the   mixed  condensate   in  OPE.
%
%
%Although we  do not include  renormalization effect, 
We note this  bare result
itself is determined very precisely~\cite{DOIS:qGq}.

%\newpage

%%%%%%%%%%%%%%%%%%%%%%%%%%%%%%%%%%%%%%%%%%%%%%%%%%%%%%%%%%%%%%%%%
%%%%%%%%%%%%%%%%%%%%%%%%%%%%%%%%%%%%%%%%%%%%%%%%%%%%%%%%%%%%%%%%%

\begin{figure}[tbh]
%\begin{minipage}[t]{155mm}
\begin{center}
\includegraphics[scale=0.3]{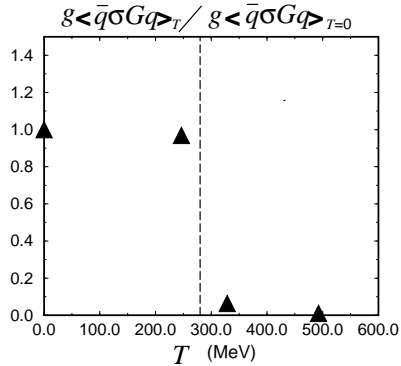}
\vspace*{-8mm}
\caption{\label{fig:qGq_finite_T}
The quark-gluon mixed condensate 
$\qGq_T/\qGq_{T=0}$
%
%$g\langle \bar q \sigma_{\mu\nu}G q (T)\rangle/g\langle \bar q \sigma_{\mu\nu}G q (T=0)\rangle$ 
%
%normalized by the zero-temperature value 
plotted against the temperature $T$.
The jackknife errors are hidden in the triangles.
The vertical dashed line denotes the critical temperature $T_c \simeq 280{\rm MeV}$ at the quenched level.
}
\end{center}
\vspace*{-8mm}
%\end{minipage}
%\hspace{\fill}
\end{figure}

%%%%%%%%%%%%%%%%%%%%%%%%%%%%%%%%%%%%%%%%%%%%%%%%%%%%%%%%%%%%%%%%%
%%%%%%%%%%%%%%%%%%%%%%%%%%%%%%%%%%%%%%%%%%%%%%%%%%%%%%%%%%%%%%%%%

\section{Discussion and Outlook}
\label{sec:summary}

For comparison with the standard value in the QCD sum rule,
we rescale our result from  $\mu\simeq\pi/a$ to $\mu\simeq 1$
GeV corresponding to the QCD sum rule.
Following Ref.~\cite{K&S},  we  first  take  the bare  values  of  the
condensates  as  the starting  point  of  the  flow, and then  rescale  the
condensates perturbatively.
We adopt the anomalous dimensions at the one-loop level~\cite{Narison2},
and choose the parameters $\Lambda_{\rm QCD} = 200-300 \mbox{MeV}$ and
$N_f=0$ corresponding to quenched lattice QCD.
We obtain
$ m_0^2  \big|_{\mu=1{\rm GeV}} \equiv  \qGq/\qq  \big|_{\mu=1{\rm GeV}} \sim  3.5-3.7 \  
{\rm GeV}^2$.
%at $\mu=1$GeV.
%
%
%from $\qq  \big|_{\mu} \sim -  (0.0477 - 0.0506 )  \ {\rm GeV}^3 
%=  - (0.36-0.37  {\rm GeV})^3$ and 
%
%$\qGq \big|_{\mu}  \sim - (0.176- 0.177)  \ {\rm GeV}^5$.
%
%
%
Comparing with the standard value of  $m_0^2 = 0.8 \pm 0.2$ GeV$^2$ in the
QCD sum rule,  our calculation results in a  rather large value. (Note
that  the instanton  model have  made  a slightly  larger estimate  as
$m_0^2 \simeq 1.4$ GeV$^2$ at $\mu\simeq 0.6$ GeV~\cite{Polyakov}.)
For the improvement, the non-perturbative
renormalization scheme may be desired.
%which is also  expected to  improve the value  of $\qq$  simultaneously.
%although this scheme would require a significant calculation cost.
% \cite{Martinelli}.

Finally, considering the importance of finite-temperature QCD in the RHIC project, 
we investigate the thermal effect on the mixed condensate $\qGq$ 
using the $16^3\times N_t$ lattices with $N_t=16, 8, 6, 4$ at $\beta=6.0$.
In figure~\ref{fig:qGq_finite_T}, we show our preliminary results for the mixed 
condensate at finite temperature.
We find a drastic change of the mixed condensate around the critical temperature $T_c$, 
which reflects the chiral-symmetry restoration.

In summary, 
we  have studied  the quark-gluon  mixed condensate  $\qGq$  using 
SU(3)$_c$ lattice QCD with  the KS-fermion at the quenched
level.
For each  quark mass of  $m_q=21, 36, 52$  MeV, we have  generated 100
gauge configurations  on the $16^4$ lattice with  $\beta =6.0$.
Using the 1600 data for each $m_q$,  
we have found $m_0^2 \equiv \qGq / \qq
\simeq 2.5$ GeV$^2$ in the chiral limit at the lattice scale corresponding to 
$\beta=6.0$ or $a^{-1} \simeq$ 2GeV.
We have also shown our preliminary results of $\qGq$ at finite temperature.

%%%%%%%%%%%%%%%%%%%%%%%%%%%%%%%%%%%%%%%%%%%%%%%%%%%%%%%%%%%%%%%%%
%%%%%%%%%%%%%%%%%%%%%%%%%%%%%%%%%%%%%%%%%%%%%%%%%%%%%%%%%%%%%%%%%


\begin{thebibliography}{9}

\bibitem{Ioffe}     {B. L. Ioffe,
                            Nucl. Phys.  {\bf B188} (1981) 317,
			    {\it Erratum-ibid.} {\bf B191} (1981) 591.}

\bibitem{Dosch}     {H.G. Dosch, M. Jamin and S. Narison,
                        Phys. Lett. {\bf B220} (1989) 251. }

\bibitem{Dosch2}     {H.G. Dosch and S. Narison,
                        Phys. Lett. {\bf B417} (1998) 173
                        and references therein.}

\bibitem{Latorre}    {J.I. Latorre, P. Pascual and S. Narison,
                        Z. Phys. {\bf C34} (1987) 347.}

\bibitem{Bel}       {V.M. Belyaev and B.L. Ioffe,
                        Sov. Phys. JETP {\bf 56} (1982) 493.}

%\bibitem{RRY2}     {L.J. Reinders, H.R. Rubinstein, and S. Yazaki,
%                        Phys. Lett. B {\bf 120}, 209 (1983).}

%\bibitem{Ovchi}     {A.A. Ovchinnikov et al.,
%                        Sov. J. Nucl. Phys. {\bf 48}, 721 (1988). }
%                        (Yad. Fiz. {\bf 48}, 1135 (1988)).}

%\bibitem{Narison1}     {S. Narison,
%                        Phys. Lett. {\bf B210}, 238 (1988).}

\bibitem{K&S}       {M. Kremer and G. Schierholz,
                        Phys. Lett. {\bf B194} (1987) 283.}

\bibitem{DOIS:qGq}  {T. Doi, N. Ishii, M. Oka and H. Suganuma,
			hep-lat/0211039 (2002).}

\bibitem{Takahashi}    {T.T. Takahashi et al.,
			Phys. Rev. {\bf D65} (2002) 114509.}

%\bibitem{Martinelli} {G. Martinelli, C. Pittori, C.T. Sachrajda,
%			M. Testa, and A. Vladikas,
%			Nucl. Phys. {\bf B445} 81 (1995).}

\bibitem{Narison2}  { S. Narison and R. Tarrach,
                        Phys. Lett. {\bf B 125} (1983) 217.}

\bibitem{Polyakov}    {M.V. Polyakov and C. Weiss,
                        Phys. Lett. {\bf B387} (1996) 841.}

\end{thebibliography}
\end{document}